\documentclass[aip,jcp,reprint,superscriptaddress]{revtex4-1}

\usepackage{color}
\usepackage{latexsym}
\usepackage{amssymb}
\usepackage{amsmath}  
\usepackage{graphicx}
\usepackage{graphics,epsfig}
\usepackage{soul}  

\def\rmd{{\mathrm{d}}}

\newcommand{\eqs}{\;\!}
\newcommand{\kB}{k_\mathrm{B}}

\newcommand{\ktot}{k_\mathrm{tot}}

\begin{document}

\title{Product interactions and feedback in diffusion-controlled reactions}

\author{Rafael Roa}
\email[]{rafaroa@uma.es}
\affiliation{\mbox{F\'isica Aplicada I, Facultad de Ciencias, Universidad de M\'alaga, 29071 M\'alaga, Spain}}
\affiliation{Institut f\"ur Weiche Materie und Funktionale Materialien, Helmholtz-Zentrum Berlin f\"ur Materialien und Energie, 14109 Berlin, Germany}
\author{Toni Siegl}
\affiliation{Institut f\"ur Weiche Materie und Funktionale Materialien, Helmholtz-Zentrum Berlin f\"ur Materialien und Energie, 14109 Berlin, Germany}
\author{Won Kyu Kim}
\affiliation{Institut f\"ur Weiche Materie und Funktionale Materialien, Helmholtz-Zentrum Berlin f\"ur Materialien und Energie, 14109 Berlin, Germany}

\author{Joachim Dzubiella}
\email[]{joachim.dzubiella@helmholtz-berlin.de}
\affiliation{Institut f\"ur Weiche Materie und Funktionale Materialien, Helmholtz-Zentrum Berlin f\"ur Materialien und Energie, 14109 Berlin, Germany}
\affiliation{Institut f\"ur Physik, Humboldt-Universit\"at zu Berlin, 12489 Berlin, Germany}

\date{\today}

\begin{abstract}

Steric or attractive interactions among reactants or between reactants and inert crowders can substantially influence the total rate of a diffusion-influenced reaction in the liquid phase. However, the role of the product species, that has typically different physical properties than the reactant species, has been disregarded so far. Here we study the effects of reactant--product and product--product interactions as well as asymmetric diffusion properties on the rate of diffusion-controlled reactions in the classical Smoluchowski-setup for chemical transformations at a perfect catalytic sphere. For this we solve the diffusion equation with appropriate boundary conditions coupled by a mean-field approach on the second virial level to account for the particle interactions. We find that all particle spatial distributions and the total rate can change significantly, depending on the diffusion and interaction properties of the accumulated products. Complex competing and self-regulating (homeostatic) or self-amplifying effects are observed for the system, leading to both decrease and increase of the rates, as the presence of interacting products feeds back to the reactant flux and thus the rate with which the products are generated.  

\end{abstract}

\maketitle

\section{Introduction}

Diffusion-controlled or diffusion-influenced reactions in the liquid phase constitute fundamental processes in physical (bio)chemistry, such as colloidal coagulation, protein association, and enzyme or nanoparticle-based catalysis, just to name a few. Since the pioneering works of Smoluchowski,~\cite{Smoluchowski:1917wh} Debye,~\cite{Debye:1942cv} and Collins and Kimball,~\cite{Collins:1949hq} a number of milestone papers have 
substantially extended and improved the theory of diffusion-influenced reactions in various directions, as summarized in seminal reviews.~\cite{Calef:1983vd, Berg:1985ea, rice:1985, Hanggi:1990en} Triggered by the more recent developments in our understanding of the biophysics of the cell as well as the  modern synthesis of complex nanoscale systems for catalysis, a stronger focus has been targeted in the last years on interaction and 
crowding effects in molecular reactions.~\cite{minton:2008, Hofling:2013bk} Reaction rates as well as the availability of reactants near 
catalytic sites are significantly modified by molecular crowding, which can be induced  either by `inert' macromolecules in the surrounding or by the reaction partners themselves. Simulations and more recent theoretical advances have shown that details are complex and depend  on the type of reaction, size and packing fraction of crowders,  and in particular the interaction types between crowders and reactants, see for instance some of the recent representative works in this direction.~\cite{Dzubiella:2005gl, weinstein:2007, benzhou, ellison, Kim:2009ky, yethiraj:2010,Dorsaz:2010hf, Zaccone2011, zaccone, Piazza2013,  Eun2013, Kekenes-Huskey2015, pete, Berezhkovskii:2016kn, Antoine:2016iu}

In that respect an important factor that has been little considered so far is that in the case of chemical molecular reactions 
{\it product species} are generated, usually right at the immediate vicinity of an enzyme or a nanoparticle that catalyzes the reaction.  
If the transformation rate is large, naturally products will locally accumulate  and may interfere with the reactants' approach for 
simple steric interaction reasons. This effect will be amplifed in crowded environments such as the biological cell,~\cite{minton:2008, Hofling:2013bk} where the products' diffusion away from the catalyst is hindered by nearby macromolecules.  
This may be additionally enhanced for enzymes or related synthetic catalysts where the catalytic sites are often 
highly local or even buried,~\cite{deutch:1978, alberty:1958, chou:1974} leading to strong confinement of the products.
That such a product accumulation might lead to a `product inhibition', that is, a negative feedback on the reaction due to steric hindrance
of the reactants by crowding products, has been realized already decades ago.~\cite{henri,pardee:1961}
However, the investigation and theoretical description of these effects has been unfortunately stopped already on a very phenomenological level within 
the standard macroscopic Michaelis--Menten framework,~\cite{horvath:1974, bowden:2013} without specifying any molecular mechanism modes for the inhibition. 

Regarding molecular interactions between reactants and products one has to realize that in general the physical properties of those can be very different. 
Consider, for instance, the redox reaction~\cite{march} of a couple R and P, where R $-$ $n{\rm e}^- \rightarrow$~P, with $n$ electrons transferred between the redox couple. In that case reactant (R) and product (P) have different electrostatic charge states and therefore different diffusion and interaction properties.  In particular, the cross-interaction may be highly non-additive, for instance, while the interaction may be repulsive between (say like-charged) particles of the same kind, it would be attractive for different particles carrying the opposite sign of charge. Other examples of many are splitting or substitution reactions changing significantly the size or polarity of a transformed molecule.~\cite{march} In fact this asymmetry between reactants and products has led to substantial works in the field of scanning electrochemical microscope and voltammetry.~\cite{unwin:1997, unwin:1998, thouin:2002, hyk:2002a, hyk:2002b} Here, experiments and theoretical treatment on the Poisson--Nernst--Planck level of the diffusion-limited transport in planar confinement driven by potential gradients showed that redox couples with {\it unequal diffusion coefficients} have dramatic effects in the transient and steady-state behavior of the currents, especially for  electrochemical detections at strongly confining nanoelectrodes.~\cite{lemay:2013}

Our work is mostly motivated by the current rapid developments in the catalysis by nanoparticles in the liquid phase.~\cite{astruc} 
Nanoparticles of various metals and metal oxides can be now well shaped and engineered, e.g., by controlled etching or the inclusion of
defects, for an optimized desired function. Furthermore they can be functionalized, e.g., by organic molecules and polymers 
to capture reactants and increase their local densities for the reaction. In particular, in an important line of developments stimuli-responsive 
shells synthesized by thermosensitive polymer architectures (brushes, networks or hydrogels) are employed, which surround the nanoparticles
or embed them.~\cite{Herves:2012fp, Wu:2012bx, AngiolettiUberti:2015go, Galanti:2016bz, Roa:2017br} The physicochemical properties of these polymeric shells can be controlled by external stimuli, such as temperature, pH, or cosolute addition, and with that the partitioning and mass 
transport of the reactants to the catalytic sites are highly selective and tunable. Polymers, reactants, and products present a quite crowded 
environment in the vicinity of the catalyst, and highly cooperative, even feedback-induced phenomena due to interactions are expected.  
A better understanding of these effects towards more rational design of those `nanoreactor' systems  might allow in future the development of programmable nanocatalysts or adaptive `colloidal enzymes'. 

The simplest models of diffusion-controlled reactions are typically described by the classical Debye--Smoluchowski setup where a central 
spherical particle acts as a sink or catalyst for molecular adsorption or transformation in a very large or infinite cell full of abundant reactants, as used in many of the previous examples.~\cite{Calef:1983vd, Berg:1985ea, rice:1985, Hanggi:1990en, Dzubiella:2005gl, Dorsaz:2010hf, AngiolettiUberti:2015go, Antoine:2016iu, Roa:2017br}  In the case of a diffusion-controlled or -limited unimolecular reaction, a reactant diffuses towards the catalytic sphere with homogeneous surface reactivity and, quickly after binding to the catalyst's surface, is transformed into a product which diffuses away from the catalyst. 
In this paper, we propose to incorporate product accumulation and interaction effects between reactants and products by employing a dynamic mean-field approximation in the diffusion equations on the second virial level. It can be formally derived as a low-density local-density approximation in a dynamic density functional framework.~\cite{marconi, hansen:book, Dzubiella:2005gl} We solve numerically the coupled equations for the time-independent (steady-state) case with appropriate boundary conditions and analyze the spatial distribution of reactants and products and the steady-state rate constant for moderate catalyst concentrations. Our analysis is made in terms of scanning different diffusion constants and interactions, the latter expressed by the second virial coefficients $B_2$ for the reactant--reactant, product--product, and reactant--product interactions. We compare our numerical results with approximate analytical solutions valid for small $B_2$ values.  We find that all particle spatial distributions and the total rate can change significantly, depending on the diffusion and interaction properties of the products. Interesting self-regulating (homeostatic) or -amplifying effects are observed, leading to both decrease and increase of the rates, as the presence of interacting products modifies the reactant flux which in turn determines the total reaction rate. 

\section{Model and Theory}

\begin{figure}[b!]
\begin{center}
\includegraphics[width=\linewidth]{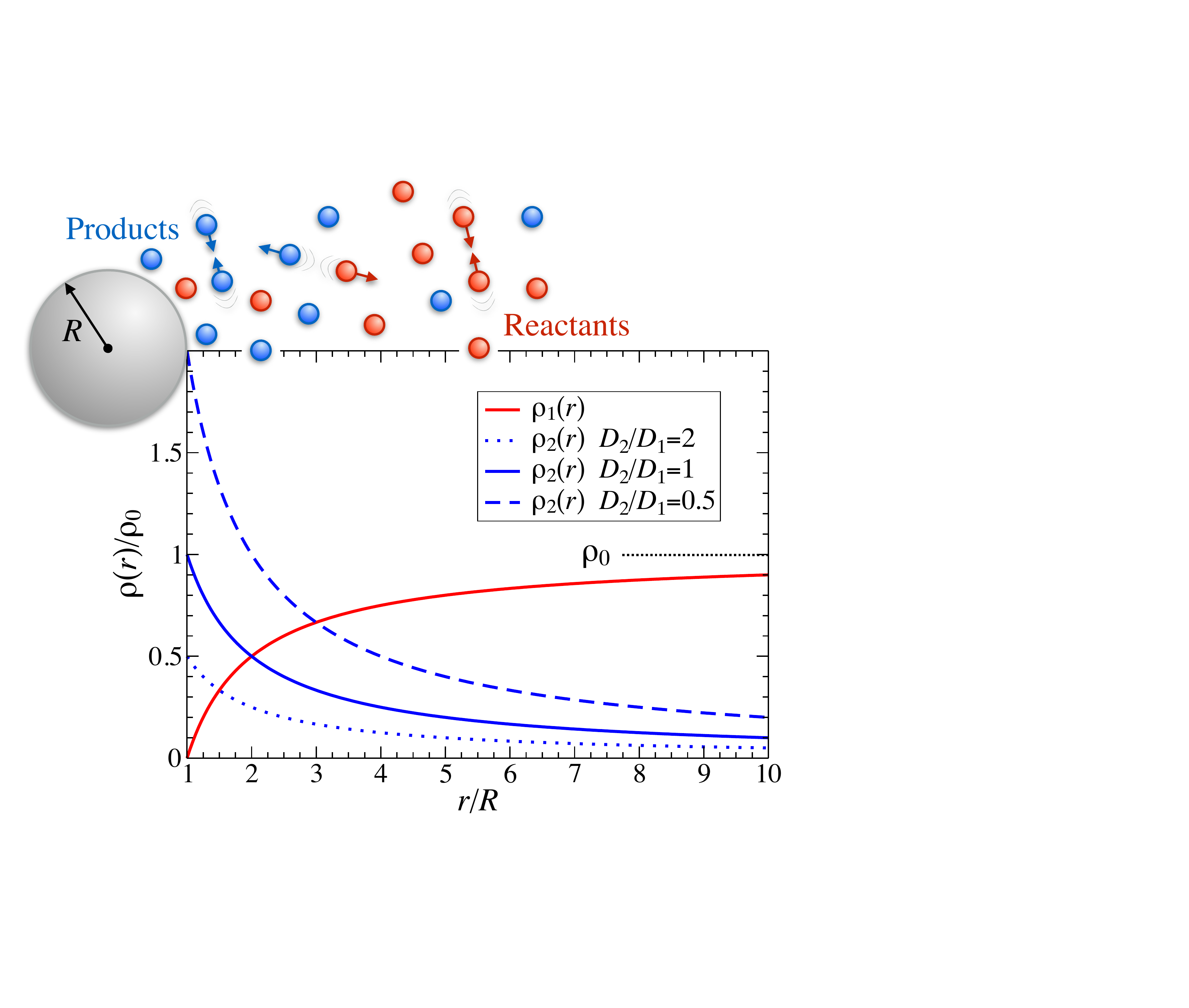}
\caption{
Sketch of the reactants (red particles) and products (blue particles) around a catalytic nanoparticle of radius $R$.
The plot shows the steady-state density profiles of reactants, $\rho_1(r)$, and products, $\rho_2(r)$, around the catalyst for different ratios of the diffusion constants, $D_2/D_1$ and no interactions, i.e., vanishing second virial coefficients $B_2^{12}=B_2^{11}=B_2^{22}=0$.
}
\label{fig_concdifu}
\end{center}
\end{figure}

We consider the classical Smoluchowski-like setup for diffusion-controlled reactions where a central spherical catalyst of radius $R$ catalyzes the conversion from reactants to products, cf. the sketch in Fig.~\ref{fig_concdifu}.  The governing equations for the dynamics of the particle densities can be derived by a popular form of classical dynamic density functional theory (DDFT)~\cite{marconi}
\begin{equation}
\frac{\partial \rho_i}{\partial t}=
\nabla\cdot
\left[
D_i(r)\rho_i(r,t)\nabla
\frac{\beta\delta\mathcal{F}[\rho]}{\delta \rho_i} 
\right]
\eqs,
\end{equation}
where $\beta=1/\kB T$, with $\kB$ denoting Boltzmann's constant, $T$ is the absolute temperature of the system, and the index $i=1,2$ labels the reactant and product particle species, respectively. The $\rho_i(r,t)$ are the particle distribution with radial distance $r$ to the catalyst center. The $D_i(r)$ are the in general position-dependent diffusion constants. We approximate the system (Helmholtz) free energy functional ${\cal F}[\rho] $ in our treatment by a local density approximation on the $B_2$-level of a virial expansion~\cite{hansen:book, Dzubiella:2005gl} 
\begin{align}\label{eq:freeenergy}
\beta\mathcal{F}[\rho]
=
\beta\mathcal{F}_\mathrm{id}[\rho]
&+
\sum_{i,j}\int \rmd^3 r \rho_i(r)\rho_j(r)B_2^{ij}\nonumber \\
&+
\sum_i \int \rmd^3 r \beta V_\mathrm{ext}^i (r) \rho_i(r)
\eqs.
\end{align}

The first term on the r.h.s. of Eq.~(\ref{eq:freeenergy}) is the ideal gas term. The second one is the excess term constituted by the local approximation of particle interactions on the $B_2$-level. The third one corresponds to the external field, in our case generated solely by the central catalytic sphere which we model as a reactive hard sphere. From the functional derivative we arrive at 
\begin{equation}
\frac{\beta\delta\mathcal{F}[\rho]}{\delta \rho_i}
=
\ln(\rho_i\Lambda_i^3)+2B_2^{ii}\rho_i+2B_2^{ij}\rho_j+\beta V_\mathrm{ext}^i (r)
\eqs.
\end{equation}

Hence, the DDFT equation yields for the time-dependent evolution of the profiles
\begin{align}
\frac{\partial \rho_i}{\partial t}=
\nabla\cdot
\bigg[
 D_i(r)\rho_i(r,t)
\bigg(
&\frac{\nabla \rho_i}{\rho_i}
+
2 B_2^{ii}\nabla\rho_i
\nonumber\\
&+
2 B_2^{ij}\nabla\rho_j
+
\nabla \beta V_\mathrm{ext}^i (r)
\bigg)
\bigg]
\eqs.
\end{align}

Assuming a constant diffusion in space in the steady-state, i.e., $\partial\rho_i/\partial t = 0$, and explicitly writing in  spherical coordinates, we finally find
\begin{align}
0
=
&\frac{\partial}{\partial r}\left(r^2\frac{\partial \rho_i}{\partial r}\right) \nonumber\\
&+
\frac{\partial}{\partial r}
\bigg[
r^2
\bigg(
2 B_2^{ii}\rho_i\frac{\partial \rho_i}{\partial r}
+
2 B_2^{ij}\rho_i\frac{\partial \rho_j}{\partial r}
+
\rho_i \frac{\partial \beta V_\mathrm{ext}^{i}}{\partial r}
\bigg)
\bigg]
\eqs. 
\end{align}
 
Therefore, the coupled equations lead to
\begin{equation}
\frac{C_1}{r^2}
=
\frac{\partial \rho_1}{\partial r}
+
\rho_1
\left(
2 B_2^{11}\frac{\partial \rho_1}{\partial r}
+
2 B_2^{12}\frac{\partial \rho_2}{\partial r}
+
\frac{\partial \beta V_\mathrm{ext}^{1}}{\partial r}
\right)
\label{eq:ddft_reactants}
\end{equation}
for the reactants, and
\begin{equation}
\frac{C_2}{r^2}
=
\frac{\partial \rho_2}{\partial r}
+
\rho_2
\left(
2 B_2^{22}\frac{\partial \rho_2}{\partial r}
+
2 B_2^{12}\frac{\partial \rho_1}{\partial r}
+
\frac{\partial \beta V_\mathrm{ext}^{2}}{\partial r}
\right)
\end{equation}
for the products.  $C_1$ and $C_2$ are integration constants that we obtain by applying the boundary conditions 
\begin{equation}
\rho_1(R)
=
0  \;\;\;
{\rm and}\;\;\;
\rho_1(\infty)=\rho_0
\end{equation}
and 
\begin{equation}
\frac{\partial\rho_2}{\partial r}\bigg{|}_{r=R}
=
-\frac{D_1}{D_2}
\frac{\partial\rho_1}{\partial r}\bigg|_{r=R} \;\;\;
{\rm and}\;\;\;
\rho_2(\infty)=0
\eqs,
\end{equation}
that is, we take the Smoluchowski limit of diffusion-controlled reactions (i.e., every reactant that arrives at the catalyst surface will be immediately transformed into a product), with opposite fluxes of reactants and products at the catalyst  surface. $D_1$ and $D_2$ are the in general different and constant diffusion coefficients of reactants and products, respectively. We also assume the infinite-dilution limit of the catalysts and with that the reactant density in bulk, $\rho_0$,  can be held constant while the products dilute away to zero far away from the catalyst. Hence, the total rate can be calculated as
\begin{equation}\label{eq:ktot}
\ktot
=
4\pi R^2 D_1
\frac{\partial\rho_1}{\partial r}\bigg|_{r=R}
\eqs. 
\end{equation}

If reactants and products do not interact with each other (or between themselves), all second virial coefficients vanish $B_2^{ij}=0$ (while higher order coefficients have been neglected anyway). Thus, with the here used hard-core potential for $V^i_{\rm ext}$, the total reaction rate becomes the classical Smoluchowski limit $k_D^0=4\pi D_1R\rho_0$.

Let us also discuss some simple analytical limits. If only the reactant--product interaction is non-zero, i.e., $B_2^{11}=B_2^{22}=0$, while $B_2^{12}\neq 0$, it is easy to show that the following analytical solutions for the steady-state reactant and product density profiles around the catalyst are valid for small values of $B_2^{12}\rho_0$, that is, low density or weak interactions,  
\begin{equation}\label{eq:rho1B212}
\rho_1(r)/\rho_0 
\approx 
1-\frac{R}{r}
\left[
1+ \frac{D_1}{D_2} B_2^{12} \rho_0 \left(1-\frac{R}{r}\right)
\right]
\eqs,
\end{equation}
\begin{equation}\label{eq:rho2B212}
\rho_2(r)/\rho_0 
\approx\frac{D_1}{D_2}\frac{R}{r}
\left[
1+ B_2^{12} \rho_0 \left(\frac{R}{r}-2-\frac{D_1}{D_2}\right)
\right]
\eqs.
\end{equation}

In this case, the total reaction rate can be approximated as
\begin{equation}\label{eq:ktotB212}
\frac{\ktot}{k_D^0}\approx 1-\frac{D_1}{D_2} B_2^{12} \rho_0
\eqs, 
\end{equation}
which demonstrates already the leading order effects of interaction and asymmetric diffusion on the total rate.  

If only the product--product interaction is non-zero, i.e., $B_2^{11}=B_2^{12}=0$, while $B_2^{22}\neq 0$, the steady-state product density profile around the catalyst takes the following analytical form valid for small  values of $B_2^{22}\rho_0$, 
\begin{equation}\label{eq:rho2B222}
\rho_2(r)/\rho_0 
\approx\frac{D_1}{D_2}\frac{R}{r}
\left[
1+\frac{D_1}{D_2} B_2^{22}\rho_0\left(2-\frac{R}{r}\right)
\right]
\eqs.
\end{equation}

To access the full solutions, we solve the coupled Eqs.~(\ref{eq:ddft_reactants})--(\ref{eq:ktot}) numerically in a self-consistent fashion. For this, a Mathematica-script was written utilizing the program-internal function NDSolve. This function also allows for the inclusion of unknown parameters (in our case the integration constants $C_1$ and $C_2$), which are returned together with the solution. The function NDSolve solves the system of coupled ordinary differential equations using the Livermore Solver for Ordinary Differential Equations.~\cite{Mathematica} 

In order to define the dimensionless interaction parameters independent from the reactant bulk density, we conveniently use the scaled values $B_2^{ij}/\sigma^3$, where $\sigma$ is the effective reactant size. The  
relation between $\rho_0$ and $\sigma^3$ is given through the definition of a reactant  
packing fraction in the bulk via $\phi_0 = \pi \rho_0\sigma^3/6$. In all following calculations, we keep the reactant bulk volume fraction fixed at $\phi_0=0.08$. This choice is somewhat arbitrary but represents a moderately dense regime where interactions play a role but is dilute enough so that the theory -- on a two-body mean-field level -- is still applicable within its approximations. In real experiments the local particle densities will depend 
very specifically on the particular system properties where they would be enhanced by local confinement, such as in buried sites or near crowding macromolecules  and/or the favorable partitioning by catalyst carrier systems. To keep our analysis on the simplest possible level 
to exemplify all the new effects, we consider the reactants to be ideal among themselves, i.e., $B_2^{11}=0$, in all cases. 
Only the influence of varying the cross-interaction, expressed by $B_2^{12}$, and product--product interactions 
$B_2^{22}$ will be investigated, as well as variations of the diffusion ratio between products and reactants, $D_2/D_1$.  

\section{Results and Discussion}

In Fig.~\ref{fig_concdifu} we first show the steady-state density profiles of reactants, $\rho_1(r)$, and products, $\rho_2(r)$, around the catalyst for different ratios of the diffusion constant, $D_2/D_1$,  while all interactions are vanishing, that is,  all interaction parameters $B_2^{12}=B_2^{11}=B_2^{22}=0$.  In this ideal limit the density profiles Eqs.~(\ref{eq:rho1B212})--(\ref{eq:rho2B212}) further simplify to 
\begin{equation}\label{eq:rho1B20}
\rho_1(r)/ \rho_0
= 
1-\frac{R}{r}, 
\end{equation}
and 
\begin{equation}\label{eq:rho2B20}
\rho_2(r)/\rho_0
=
\frac{D_1}{D_2}\frac{R}{r}. 
\end{equation}

We make the following interesting observations. First, and as expected, in this non-interacting case the reactant density profile does not depend on the diffusion coefficient of reactants and products. However, the product density depends on the ratio of the diffusion constants. In the symmetric limit,  $D_2=D_1$, the product density at the catalyst surface equals the bulk density of reactants, $\rho_2(R)/\rho_0=1$.  Also, the sum of the local concentrations of reactants and products $\rho_1(r)+\rho_2(r)=\rho_0$ everywhere. If $D_2>D_1$, then $\rho_2(R)/\rho_0<1$. In this case, the sum of the local concentrations of reactants and products $\rho_1(r)+\rho_2(r)< \rho_0$ and is different at every distance $r$. Vice versa, if $D_2<D_1$, then $\rho_2(R)/\rho_0>1$ and 
$\rho_1(r)+\rho_2(r)>\rho_0$ is not constant. 
The latter scenario gives rise to an accumulation of products at the vicinity of the catalyst which can be regarded also in some sense as a `product crowding'.
Hence, we conclude from this simple example that already different mobilities of products and reactants lead to structural redistributions in the steady-state profiles which will naturally have influences on the fluxes and rate in interacting systems, as already known for currents of redox couples at nanoelectrodes.~\cite{lemay:2013}

\begin{figure}[t!]
\begin{center}
\includegraphics[width=\linewidth]{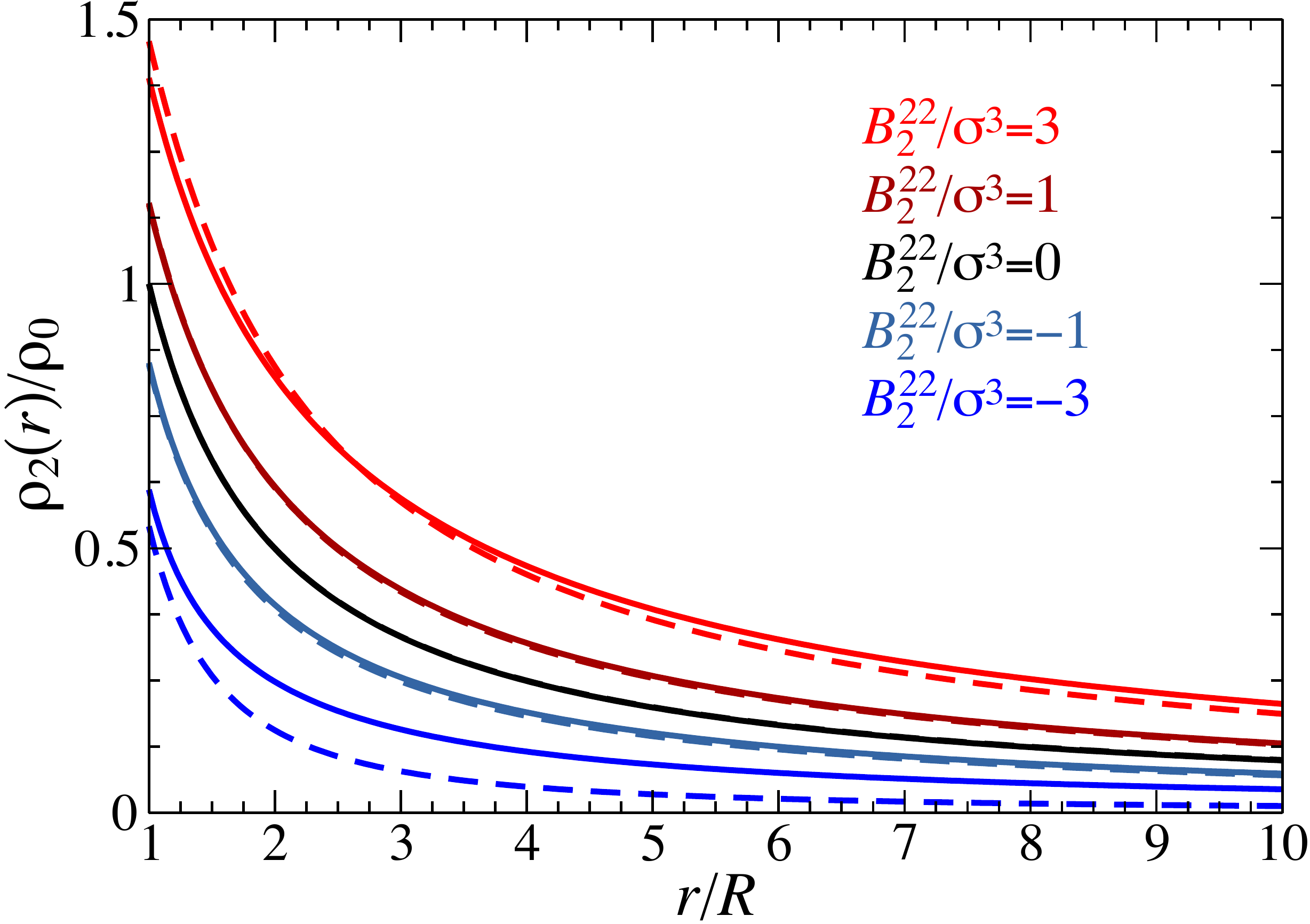}
\caption{Steady-state density profiles of the products, $\rho_2(r)$, around the catalyst for different $B_2^{22}$ values (different solid colored lines) with $B_2^{11}=B_2^{12}=0$ and $D_2=D_1$. The dashed lines stand for the approximate analytical solutions given by Eq.~(\ref{eq:rho2B222}).
}
\label{fig_concB222}
\end{center}
\end{figure}

In the next step we present in Fig.~\ref{fig_concB222} the steady-state density profiles of products, $\rho_2(r)$, around the catalyst for the case of different product--product interaction $B_2^{22}$ values (different solid colored lines) while keeping $B_2^{11}=B_2^{12}=0$ and $D_2=D_1$. Hence, only product--product interactions are imposed while reactant self- and cross-interactions are ideal. Here, for attractive (repulsive) product--product interactions $B_2^{22}<0$ ($B_2^{22}>0$), the product concentration in the vicinity of the catalyst is reduced (increased) with respect to the non-interacting case ($B_2^{22}=0$).  
The analytical result for the concentration at the catalyst surface ($r=R$) given by Eq.~(\ref{eq:rho2B222}), valid for small $B_2^{22}/\sigma^3$ values, simplifies in this case to a simple virial-like expansion $\rho_2(R)/\rho_0 = 1 + B_2^{22} \rho_0$ and qualitatively agrees with these trends.
Hence, this behavior can be explained by a simple equilibrium picture where the increasing (or decreasing) bulk osmotic pressure pushes (draws) particles from the solid surface, as in simple attractive fluids near a hard wall.~\cite{hansen:book}
Therefore, if the products increasingly attract (repel) each other, they will be more depleted (more accumulated) at the vicinity of the catalyst.

\begin{figure}[t!]
\begin{center}
\includegraphics[width=\linewidth]{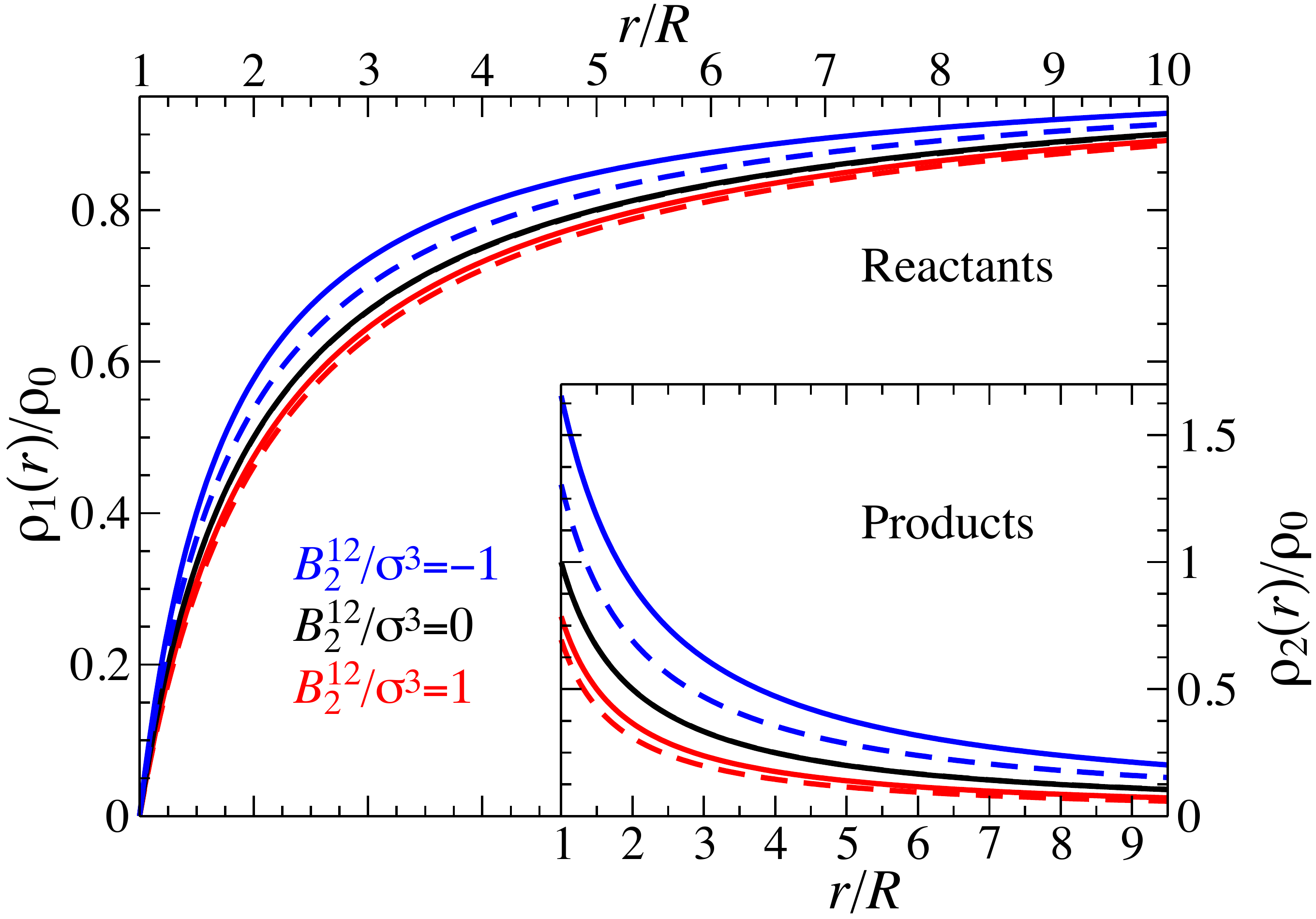}
\caption{Steady-state density profiles of reactants, $\rho_1(r)$,  and products, $\rho_2(r)$, (inset) around the catalyst for different $B_2^{12}$ values (different solid colored lines) with $B_2^{11}=B_2^{22}=0$ and $D_2=D_1$. The dashed lines stand for the approximate analytical solutions for reactants and products given by Eqs.~(\ref{eq:rho1B212}) and (\ref{eq:rho2B212}), respectively.
They agree better with the numerical solutions for repulsive reactant--product interactions, $B_2^{12}>0$, than for attractive reactant--product interactions, $B_2^{12}<0$. }
\label{fig_concB212}
\end{center}
\end{figure}

In Fig.~\ref{fig_concB212} we now show the steady-state density profiles of reactants, $\rho_1(r)$, around the catalyst for different reactant--product  interactions, expressed by scanning $B_2^{12}$ values (different solid colored lines) keeping $B_2^{11}=B_2^{22}=0$ and $D_2=D_1$. 
In other words, the system is symmetric and self-interactions are absent, only a cross-interaction between reactants and products is established.  The corresponding product profiles, $\rho_2(r)$, are displayed in the inset  to Fig.~\ref{fig_concB212}.  
We find that for attractive (repulsive) reactant--product interactions $B_2^{12}<0$ ($B_2^{12}>0$), the reactant concentration in the vicinity of the catalyst is increased (reduced) with respect to the non-interacting case ($B_2^{12}=0$). The reason is that the naturally large density of products in the vicinity of the catalyst overall attracts ($B_2^{12}<0$) or repels ($B_2^{12}>0$) the reactant distribution. This will consequently lead to an increase (or decrease) of the production rate, as shown explicitly further below.  In turn, for attractive (repulsive) reactant--product interactions thus also the product concentration in the vicinity of the catalyst is increased (reduced) with respect to the non-interacting case ($B_2^{12}=0$).  Hence, the ubiquitous presence of crowding or attracting products not only changes the reactant distributions (like inert crowders, e.g.,~\cite{weinstein:2007, yethiraj:2010, Dorsaz:2010hf}) but necessarily has interesting feedback effects, coupling product generation back to the reactant flux. This leads to homeostatic, i.e., self-regulating effects for repulsive cross-interaction (due to negative feedback) and self-amplification for attractive cross-interactions (due to positive feedback), as can be seen from the much bigger effects for the $B_2^{12}<0$ case (blue lines). 

\begin{figure}[t!]
\begin{center}
\includegraphics[width=\linewidth]{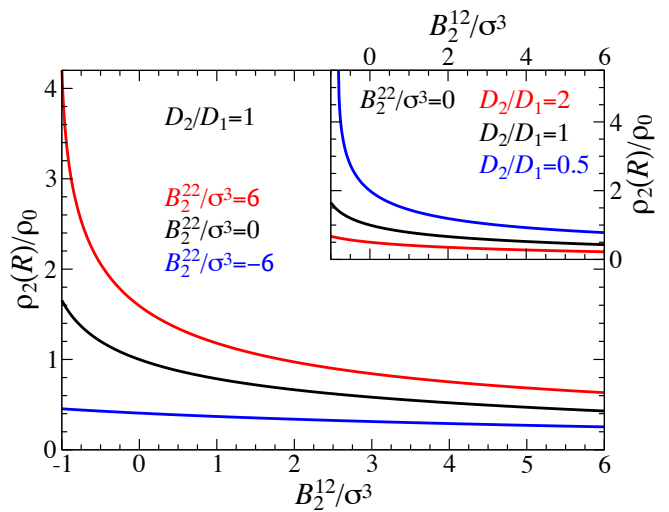}
\caption{
Steady-state product concentration at the catalyst surface, $\rho_2(r=R)$, as a function of the reactant--product interaction parameter, $B_2^{12}$, for different $B_2^{22}$ values with $D_2=D_1$, and $B_2^{11}=0$. Inset: Again $\rho_2(r=R)$ is plotted versus $B_2^{12}$ but now for different $D_2/D_1$ values with $B_2^{11}=B_2^{22}=0$.}
\label{fig_rho2R}
\end{center}
\end{figure}

\begin{figure}[t!]
\begin{center}
\includegraphics[width=\linewidth]{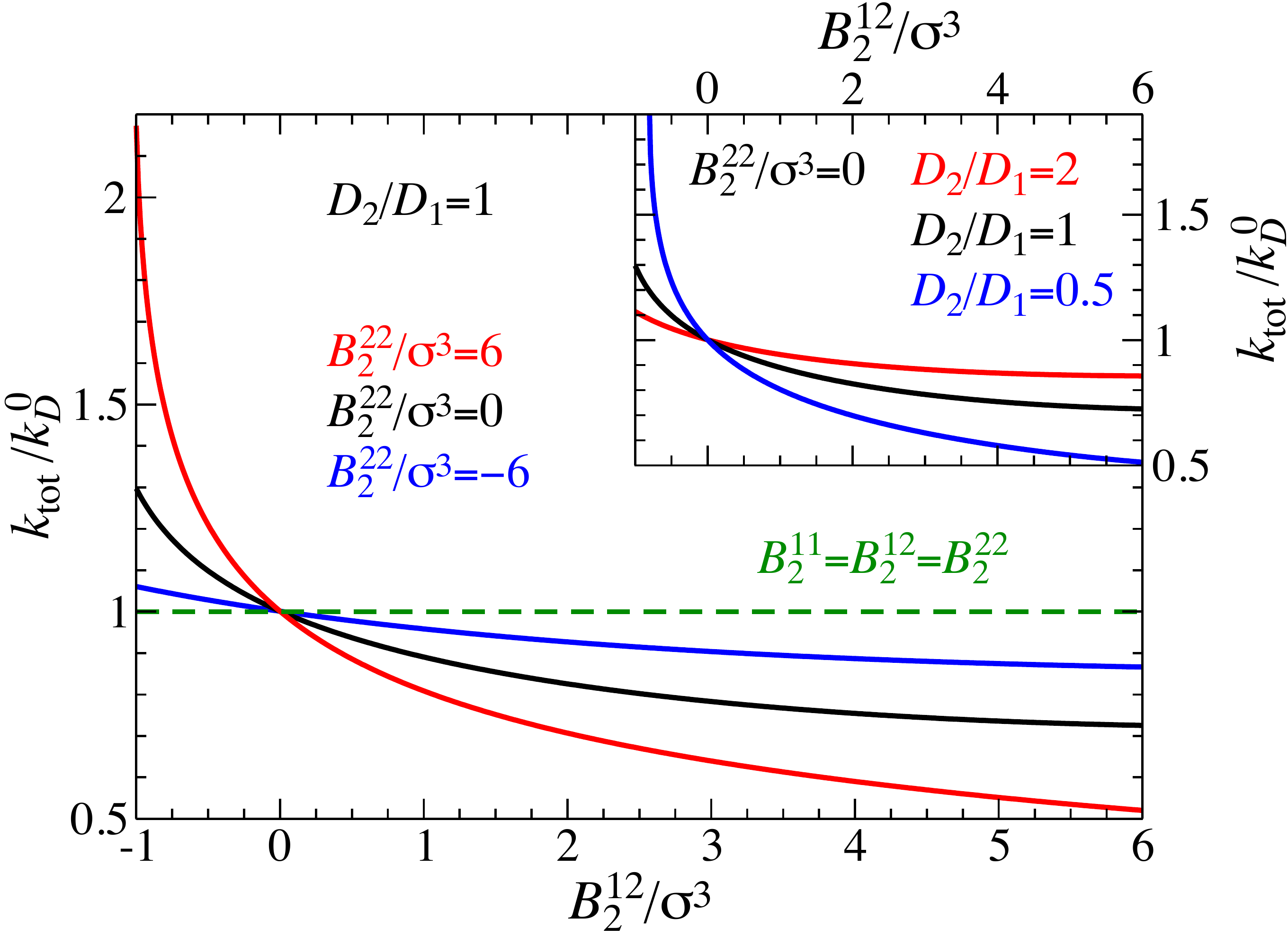}
\caption{
Total reaction rate normalized by the Smoluchowski rate, $k_\mathrm{tot}/k_D^0$, as a function of the reactant--product interaction parameter, $B_2^{12}$, for different $B_2^{22}$ values with $D_2=D_1$, and $B_2^{11}=0$. 
The dashed green line is the solution for the fully symmetric case of equal virial coefficients $B_2^{11}=B_2^{12}=B_2^{22}$.
Inset: Again $k_\mathrm{tot}/k_D^0$ is plotted versus $B_2^{12}$ but now for different $D_2/D_1$ values with $B_2^{11}=B_2^{22}=0$.
}
\label{fig_rate}
\end{center}
\end{figure}

In Fig.~\ref{fig_rho2R} we combine the results from Figs.~\ref{fig_concB222} and~\ref{fig_concB212} to present the steady-state product density {\it right at the catalyst surface},  $r=R$, and extend them to a larger range of reactant--product and product--product interactions. 
As we see, the product concentration (implicitly reflecting the qualitative behavior of the production rate, see further below) can be tuned by the various choices of interaction parameters and diffusion constants by one order of magnitude, ranging from 1/5 of the reactant bulk concentration up to 5 times the bulk concentration.   

This brings us to the discussion of the influence of the interaction and diffusion properties on the total reaction rates: In Fig.~\ref{fig_rate} we show the total reaction rate normalized by the ideal Smoluchowski rate, $k_\mathrm{tot}/k_D^0$, as a function of the reactant--product interaction parameter, $B_2^{12}$, for different $B_2^{22}$ values with $D_2=D_1$. Taking the ideal limit $k_\mathrm{tot}/k_D^0=1$ as a reference, we observe that for all $B_2^{22}$ values the total rate decreases (increases) for increasing reactant--product repulsion $B_2^{12}>0$ (attraction $B_2^{12}<0$).  This is in accordance with what we observed in the density profiles in Fig.~\ref{fig_concB212}: The flux of reactants at the catalyst surface increases (decreases) for attractive (repulsive) $B_2^{12}$ values. This effect is enlarged (diminished) if the products repel $B_2^{22}>0$ (attract $B_2^{22}<0$) each other, consistent with the product density trends close to the catalyst observed in Figs.~\ref{fig_concB222} and \ref{fig_rho2R}: If $B_2^{22}>0$ ($B_2^{22}<0$), the product  density in the vicinity of the catalyst increases (decreases) with respect to the $B_2^{22}=0$ case, leading to an amplification (reduction) of the effects of the reactant--product interaction in comparison with the $B_2^{22}=0$ case. We emphasize again the asymmetry in the cross-interaction effects: The attraction $B_2^{12}/\sigma^3= -1$ leads to much larger effects due to self-amplification in the system, while $B_2^{12}/\sigma^3= +1$ acts much smaller due to negative feedback. 
Note that if all the interaction parameters have the same value ($B_2^{11}=B_2^{12}=B_2^{22}$, dashed green line) for symmetric reactants and products, the total reaction rate retains the ideal Smoluchowski rate value.

Finally, in the inset to Fig.~\ref{fig_rate} we present the normalized total reaction rate again as a function of $B_2^{12}$ but for different $D_2/D_1$ values with $B_2^{22}=0$, i.e., only cross-interactions are non-vanishing.  Recall that, according to Fig.~\ref{fig_concdifu}, if $D_2<D_1$ ($D_2>D_1$), the product concentration in the vicinity of the catalyst increases (decreases) with respect to the symmetric $D_2=D_1$ case.  If $D_2<D_1$, the results are qualitatively similar than the ones for $B_2^{22}>0$. This is as expected, since in both cases we increase the product density in the vicinity of the catalyst, leading to an enhanced effect on the approaching reactants. Analogously, if $D_2>D_1$, the results are similar than the ones for $B_2^{22}<0$. 

\section{Concluding remarks}

In summary, we have studied the effect of the ubiquitous presence of locally accumulated interacting products on 
the spatial distribution of reactants and products and the final rate of diffusion-controlled reactions at a central catalyst, using a simple form of a mean-field dynamic density functional theory. We considered implicitly in our treatment 
that reactants and  products may have different physical properties (size, charge, etc.) which in general results into 
different diffusion and interaction behaviors, in particular exhibiting non-additive cross-interactions between reactants and products 
which we quantified by tunable second virial coefficients. Despite the simplicity of our model, we found complex and intriguing results due to the dynamical coupling of the particles fluxes by the interactions. Reaction rates and particle densities vicinal to the catalyst were found
to significantly decrease or increase, depending on the combination of parameters. Most importantly, repulsive cross-interaction 
leads to product crowding effects accompanied by negative feedback and self-regulation of the rate. Attractive cross-interaction, 
on the other side, leads to positive feedback and the amplification of rates. 

Related feedback-controlled  scenarios are already known for enzymes, where reaction products can promote or inhibit the very reaction, for instance by 
allosteric back-associating with enzymes within the complex catalytic chain.~\cite{pardee:1961} Our scenario is a bit different as it is based on simple particle--particle interactions and applies in the case of local product accumulation whose effects up to now have been only very phenomenologically discussed.~\cite{horvath:1974, bowden:2013} These effects are expected to be strong for products that find themselves locally confined after the reaction, for instance in buried catalytic sites or in the presence of crowders. The feedback by accumulated products to the reaction could be very relevant in the emerging field of nanoparticle catalysis where for instance metallic nanoparticles catalyze oxidation or reduction reactions of small molecules.~\cite{astruc,Herves:2012fp, Wu:2012bx}  We note that our virial formalism can be easily incorporated in transport equations dealing with less symmetrical problems including explicitly electrostatics, e.g.,~\cite{benzhou} bimolecular reactions~\cite{Roa:2017br} with or without interacting intermediate species, as well as with more complex boundary conditions and local external potentials that define the local partitioning of particles.

\begin{acknowledgments}
The authors thank Matej Kandu\v{c}, Gregor Wei{\ss}, Yi-Chen Lin, and Matthias Ballauff for useful discussions.
This project has received funding from the European Research Council (ERC) under the European Union's Horizon 2020 research and innovation program (grant agreement no. 646659-NANOREACTOR).
\end{acknowledgments}


\bibliographystyle{aipnum4-1}
\bibliography{literature}

\end{document}